\documentclass[
 preprint,
superscriptaddress,
longbibliography,
 amsmath,amssymb,
 aps,
]{revtex4-1}
\setcitestyle{numbers,square}

\usepackage{braket}
\usepackage{graphicx}% Include figure files
\usepackage{dcolumn}% Align table columns on decimal point
\usepackage{bm}% bold math
\usepackage[normalem]{ulem} % strikethrough
\usepackage{xcolor}
\usepackage{hyperref}% add hypertext capabilities
\usepackage{siunitx}
\usepackage{mathtools}
\usepackage{soul} % to use \st for strikeout 
\usepackage[normalem]{ulem} % to use \sout for strikeout to include equations

\hypersetup{
    colorlinks,%
    citecolor=blue,%
    linkcolor=blue,%
    urlcolor=blue
}

\newcommand{\orcid}[1]{\href{https://orcid.org/#1}{\includegraphics[width=8pt]{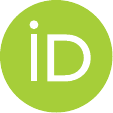}}}
\begin{document}

\title{Engineering Quantum Phases in Two Dimensions via Vacancy-Induced Electronic Reconstruction}

\author{E. V. C. Lopes \orcid{0000-0002-7981-2161}
}
\email{emmanuel.lopes@ilum.cnpem.br}
\affiliation{Ilum School of Science, Brazilian Center for Research in Energy and Materials (CNPEM), Campinas, SP, Brazil.}

\author{F. Crasto de Lima \orcid{0000-0002-2937-2620}
} 
\email{felipe.lima@ilum.cnpem.br}
\affiliation{Ilum School of Science, Brazilian Center for Research in Energy and Materials (CNPEM), Campinas, SP, Brazil.}

\author{Caio Lewenkopf \orcid{0000-0002-2053-2798}}
\email{lewenkopf@if.ufrj.br}
\affiliation{Instituto de F\'{\i}sica, Universidade Federal do Rio de Janeiro, 21941-972 Rio de Janeiro, RJ, Brazil}

\author{A. Fazzio \orcid{0000-0001-5384-7676}}
\email{adalberto.fazzio@ilum.cnpem.br}
\affiliation{Ilum School of Science, Brazilian Center for Research in Energy and Materials (CNPEM), Campinas, SP, Brazil.}
\date{\today}

%%%%%%%%%%%%%%%%   ABSTRACT 
\begin{abstract}
Topological phases of matter are commonly understood as emerging either from crystalline symmetry and intrinsic spin–orbit coupling or from disorder-driven electronic renormalization. In realistic materials, however, structural defects naturally combine both ingredients. Here, we demonstrate a general and material-independent mechanism by which atomic vacancies can induce topological phase transitions in two-dimensional semiconductors that are otherwise topologically trivial. Vacancies generate locally ordered dangling-bond states governed by well-defined hopping and spin–orbit interactions, while their spatial distribution and mutual coupling introduce long-range disorder. As vacancy concentration increases, the hybridization of these defect states forms an emergent electronic subspace that undergoes a topological transition. Using a tight-binding framework supported by large-scale density functional theory calculations, we show that this vacancy-induced electronic reconstruction can robustly stabilize quantum spin Hall, quantum anomalous Hall, and Weyl semimetal phases, depending on symmetry breaking and spin polarization. Our results establish vacancies not merely as perturbations, but as active design elements capable of transforming trivial insulators into topological quantum matter, opening realistic routes for defect-engineered topological devices.

\end{abstract}

\maketitle

The architecture of modern electronic devices is fundamentally built upon the precise engineering of semiconductor materials.
Historically, native point defects -- such as substitutional impurities and vacancies -- were viewed deleterious imperfections, that limited reproducibility and degraded carrier mobility and optical performance \cite{PRshockley1952, PRBkohan2000}.
However, a transformative paradigm shift has reclassified these defects as active degrees of freedom that enable device functionality.
Rather than limiting performance, vacancy migration supports resistive switching \cite{NATNANOyang2008, PRLzhang1991}, defect states can enhance photoluminescence \cite{ACSNANOnan2014}, and charge transport in low-dimensional semiconductors \cite{NATCOMMqiu2013}.
By treating vacancies as intentional design elements rather than stochastic errors, it is now possible to induce emergent magnetic order \cite{PRBmosca2023, JPDAParaujo2022} and achieve effective routes for doping and carrier control \cite{PRLkomsa2012}.
This approach paves the way for a new generation of semiconductor platforms that integrate functional electronic and spintronic architectures.

In parallel, the discovery of topological phases of matter, most notably topological insulators (TIs) \cite{PRLkane2005, SCIENCEbernevig2006, NATPHYSxia2009, RMPQi2010, RMPhasan2010}, has introduced a fundamentally new paradigm for electronic functionality.
Topological phases host robust boundary states protected by global symmetries, such as time-reversal symmetry, offering a platform for dissipationless transport. 
While strong disorder can degrade or even destroy these phases \cite{PRLagarwala2017, PRBfocassio2021},
the inverse phenomenon -- where disorder itself drives topology -- has  emerged as a powerful counter-intuitive effect.
In topological Anderson insulators \cite{PRLli2009, PRLgroth2009, PRBL-Assuncao2024} and amorphous-driven topological insulators  \cite{PRBL-Regis2024}, a trivial band structure becomes topologically nontrivial due to disorder-driven renormalization of electronic states.
Such disorder-induced phases open promising frontiers for low-dissipation electronics \cite{Yan2024}, spin–orbit torque devices \cite{SSCedelstein1990, NATMATfan2014, NATCOMMwang2017}, magnetic memory \cite{NATmanipatruni2018}, sensing \cite{Li2022, Xiao2023}, and quantum information technologies \cite{PhysRevB.96.115407, Ohfuchi2023}.

From a theoretical perspective, two complementary routes have traditionally defined our understanding of topological phases. 
Ordered lattice models, such as the Kane–Mele model \cite{PRLkane2005}, capture the emergence of topology from intrinsic spin–orbit coupling and local crystal symmetry.
Conversely disordered models, exemplified by topological Anderson insulators \cite{PRLli2009, PRLgroth2009}, highlight the constructive role of randomness and long-range disorder.
In realistic materials, however, defects, particularly vacancies, 
represent a unique duality: they combine local structural order with global spatial disorder. 
A vacancy locally preserves the symmetry of its crystalline environment, yielding a well-defined set of dangling-bond states (DBS), while their collective spatial distribution are inherently disordered and sample-dependent. 
Despite the prevalence of such defects, a unified framework 
{that bridges microscopic} defect physics with 
{macroscopic}
topological phase transitions in realistic materials remains largely unexplored.

Here we show a general, material-independent mechanism by which 
engineered vacancy concentrations drive topological phase transitions in otherwise trivial semiconductors. 
This transition arises from the interplay between locally ordered electronic structure, governed by intra-vacancy hopping and spin–orbit coupling in a Kane–Mele–like fashion, and long-range disordered coupling between vacancies.
We demonstrate that as the vacancy concentration increases, initially localized defect states hybridize to form an emergent electronic subspace that undergoes a topological transition. 
{Notably, topological protection stabilizes this phase against a wide range of random spatial vacancy distributions, and the critical concentrations required for the transition remain well within current experimental standards for defect engineering.}
Using effective tight-binding models validated by large-scale first-principles calculations, we establish that this mechanism robustly generates quantum spin Hall (QSH), quantum anomalous Hall (QAH), and Weyl semimetal (WSM) phases.
The resulting topological state is determined by a controllable hierarchy of vacancy density, symmetry breaking, and spin polarization, providing a universal blueprint for defect-induced topology.

\section*{Vacancy-Induced Topological Phase Mechanism}

\begin{figure}[h!]
    \centering
    \includegraphics[width=0.6\linewidth]{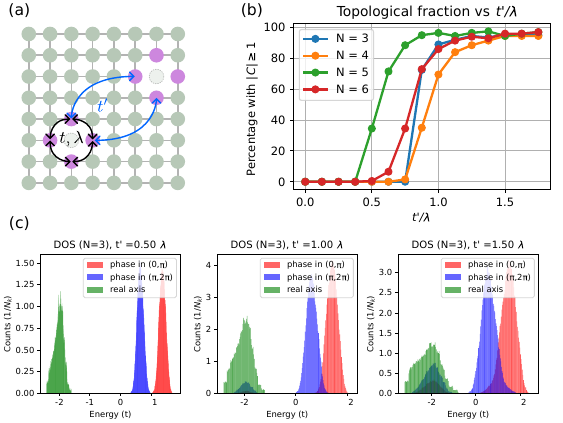}
    \caption{\label{fig:vacancy-driven} 
    {\bf Vacancy interaction model and topological phase transition.} 
    (a) Schematic of vacancy local environment, where the DBS are highlighted in purple. 
    {Arrows indicate the intra-vacancy hopping $t$ and spin-orbit coupling $\lambda$ within the ordered local host, alongside the inter-vacancy coupling $t'$ that governs long-range interactions. }
    (b) Fraction of systems presenting non-trivial topology, where $N$ represents the number of sites for each vacancy environment. 
    (c) Eigenstates evolution for increasing value of inter-vacancy interaction. The DOS plot for all the 1000 systems together, with $N_k$ the total number of k-points. The color os the DOS states are related to the accumulated phase over a cyclic evolution in the vacancy environment (see Methods).}
\end{figure}

{
The formation of a vacancy in a crystal reshapes the local electronic environment, generating a discrete set of DBS from the neighboring atomic orbitals, see Fig. \ref{fig:vacancy-driven}a. 
Our {\it ab initio} simulations (discussed in detail below) reveal that these vacancy states frequently form an isolated electronic subspace within the energy gap of a wide range of two-dimensional (2D) semiconductors.
}

{
To establish the microscopic origin of the topological potential of these states, we first consider the microscopic interactions governing the vacancy site. 
At the local scale, the DBS are spatially constrained by the point-group symmetry of the crystalline host. 
This symmetry dictates a structured intra-vacancy interaction—comprising both a kinetic hopping term $t$ and a local spin-orbit coupling $\lambda$, as detailed in the Methods. 
Furthermore, depending on the electronic occupation and exchange-correlation effects, these defect sites may host a net magnetization, potentially breaking time-reversal symmetry.
}

{
In a trivial insulator, an isolated vacancy remains a localized perturbation that does not alter the global topology.
That is, a strictly localized state corresponds to a vanishing Chern number \cite{NPcaio2019}.
However, as the vacancy concentration increases, the inter-vacancy coupling, denoted by the hopping parameter $t'$ in Fig.~\ref{fig:vacancy-driven}(a), facilitates the hybridization of these states. 
This coupling transforms the discrete DBS into a dispersive defect band.
Because the inter-vacancy interaction is sensitive to the spatial distribution of the defects, the resulting electronic structure represents a competition between local symmetry and global disorder.
We thus model the system as a dual-natured landscape: a locally ordered interaction (governed by a Kane–Mele-type Hamiltonian) superimposed on a long-range disordered network.
This framework is particularly relevant given that vacancies in transition metal dichalcogenides (TMDs) can be realized both as random distributions \cite{NATCOMMqiu2013, PRLkomsa2012, AMTneves2025} or as ordered superlattices driven by thermal annealing \cite{NATMATlin2017, NATNANOfang2023}.
}

%========================  Figure 1  ==========================
%========================  Figure 1  ==========================

Guided by these arguments, we construct {an effective} Hamiltonian {describing} $N$-locally states, {corresponding to the DBS of a single vacancy, subject to a stochastic (adjacent) inter-vacancy coupling (see Methods).}
{To ensure statistical significance, we perform an ensemble analysis over 1,000 random configurations for each local environment geometry.}
{By computing the Chern number $C$ for each realization, we map the probability of topological emergence as a function of the dimensionless ratio $t'/\lambda$, representing the competition between inter-vacancy hybridization and local spin-orbit strength.} 
{
As shown in Fig.~\ref{fig:vacancy-driven}(b), the fraction of topologically non-trivial configurations increases sharply for $t'/\lambda > 0.75$, signaling a robust phase transition. 
This transition is further characterized by Fig.~\ref{fig:vacancy-driven}(c), which displays the evolution of the density of states (DOS) for a triangular local environment ($N=3$), here we show the combined density of states for the 1000 systems, where $N_k$ is the total number of k-points. 
The ensemble-averaged spectrum reveals a clear band inversion, driven by the overlap of states with distinct phase accumulation (defined in Methods). That is, there are states with a phase accumulated on a cyclic evolution in the vacancy local environment between $(0,\pi)$ and $(\pi, 2\pi)$. With the increase of inter-vacancy interaction, this band inversion ultimately giving rise to a well-defined topological gap.}

%=======================  Figure 2 ==========================
\begin{figure}[h!]
    \centering
    \includegraphics[width=0.8\linewidth]{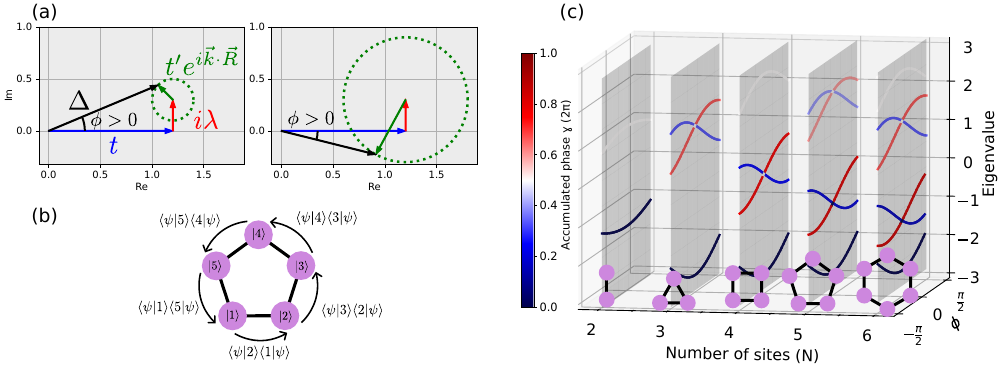}
    \caption{\label{fig:vacancy-phase} {\bf Topological phase interpretation} (a) Hamiltonian terms for lower (left) and higher (right) inter-vacancy interaction $t'$. (b) Accumulate phase over cyclic evolution. (c) Eigenstates and respective accumulated phase (color) for different local vacancy environment. }
\end{figure}
%=======================  Figure 2 ==========================

{These findings strongly suggest that the specific spatial arrangement of vacancies is secondary to the fundamental electronic mechanism driving the transition.
Therefore we are entitled to simplify the model. 
To get further insight, let us assume a simple complex coupling amplitude $\Delta = t \pm i\lambda + t' e^{i\vec{k} \cdot \vec{R}}$, for all DBS.}
This interaction, 
{which encapsulates intra-vacancy hopping $t$, spin–orbit coupling $\lambda$, and inter-vacancy hybridization $t'$} can be represented by a vector in the complex plane, $\Delta = \vert \Delta \vert e^{i \phi}$, see Fig.~\ref{fig:vacancy-phase}(a).
The accumulated phase over the cyclic evolution in the vacancy sites, Fig.~\ref{fig:vacancy-phase}(b), and the system eigenvalues Fig.~\ref{fig:vacancy-phase}(c), are related to the angle between $\Delta$ and the real axis, see Methods section.
There is a double degeneracy point for each spin subspace for $\phi=0$, that is $\Delta$ real.
A gap between those degenerated states appears for $\phi \neq 0$, while there is not an adiabatic connection preserving this gap connecting the $\phi<0$ and $\phi>0$ phase.
Notice that this comes from the accumulated phase of the cyclic evolution that is different for those eigenstates, Fig.~\ref{fig:vacancy-phase}(c).
With this interpretation, and the vector analysis of Fig.~\ref{fig:vacancy-phase}(a), we can see that the spin-orbit-coupling takes the system towards $\phi>0$, while increasing the inter-vacancy interaction term $t'$, allow the system to map also the $\phi<0$ space, that is, leading to a gap inversion.

The analysis above establishes a {universal,} material-independent criterion for vacancy-driven topology: a locally ordered set of dangling-bond states endowed with spin–orbit coupling, whose collective hybridization is tuned by the inter-vacancy hopping scale.
{Remarkably, the emergence of topology is robust against spatial disorder,}
provided that defect states remain energetically isolated from the {bulk} bands and sufficiently extended to facilitate hybridization.
In realistic materials, these requirements translate into a set of microscopic conditions: 
(i) a sizable pristine band gap to host an isolated defect subspace, 
(ii) heavy atomic constituents to provide the necessary spin-orbit strength, and 
(iii) vacancy-induced orbitals with enough spatial extent to enable tunable inter-vacancy coupling as the experimental defect densities.
{By mapping these criteria onto first-principles calculations, we can transition from abstract models to the systematic search of topological phases in real two-dimensional semiconductors.}

%%%%%%%%%%%%%%%%%%%%%%%%%%%%%%%%%%%%%%%%%%%%%%%%%%%%%%%%%%%%%%%%%%%%%
\section*{Material Realizations of Vacancy-Induced Topology}

\begin{figure*}[h!]
\includegraphics[width=\columnwidth, height=5.0cm]{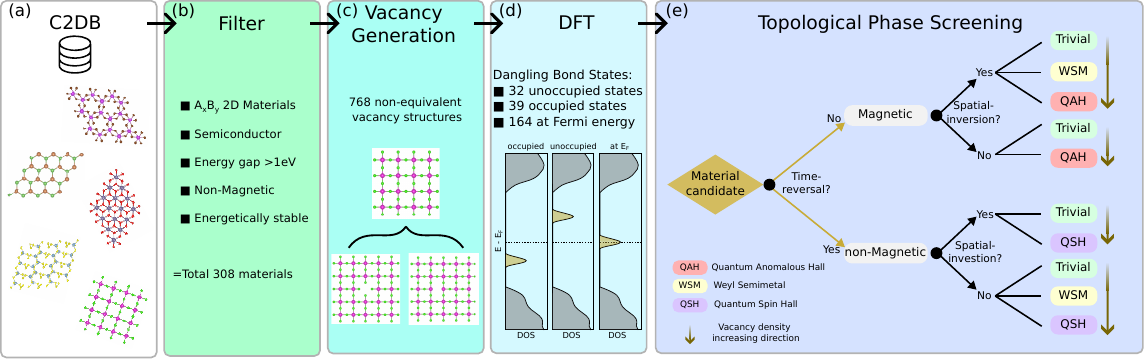}
\caption{\label{fig:painel} 
\textbf{High-throughput screening and topological phase classification.} 
{
(a) Computational workflow for the identification of candidate 2D semiconductors from the C2DB database. 
(b) Selection criteria used to filter the database, including pristine bandgap, magnetic ground state, and thermodynamic stability. 
This filtering identified 308 host materials, which were systematically expanded into 768 symmetry-unique vacancy configurations, as shown in (c).
(d) Distribution of the vacancy-induced states relative to the host bandgap; out of the 235 systems with isolated defect bands, 164 exhibit a density of states  pinned at the Fermi level. 
(e) Schematic representation of the universal topological phase screening. The trajectories illustrate the emergence of Weyl semimetal (WSM), quantum anomalous Hall (QAH), and quantum spin Hall (QSH) phases as a function of vacancy density and symmetry-breaking parameters.
}
}
\end{figure*}

{We performed a high-throughput screening} for candidate systems using the Computational 2D Materials Database (C2DB) \cite{c2db}.
From this set, we selected binary semiconductors that satisfy three stability and electronic criteria: 
(i) a pristine band gap exceeding $1.0$\,eV to ensure that vacancy-induced states remain energetically separated from the 
{bulk} 
bands; 
(ii) a non-magnetic ground state in the pristine phase; 
and (iii) {a dynamic and thermodynamic stability defined} by an energy above the convex hull of less than $0.1$\,eV. 
This multi-step filtering procedure, illustrated in Figs.~\ref{fig:painel}(a) and \ref{fig:painel}(b), 
{was designed to isolate materials where vacancy physics can be probed without interference from intrinsic metallic or magnetic instabilities.}
For the filtered candidates, we 
{systematically}
generated all symmetry-nonequivalent vacancy configurations. 
{To initially} minimize vacancy–vacancy interactions within the supercell framework, we enforced a minimum separation of $10$\,{\AA} between adjacent vacancies. 
{This rigorous down-selection process reduced the initial database to a refined set of $308$ host materials} with $768$ distinct, structurally stable vacancy configurations [Fig. \ref{fig:painel}(c)].

Our {\it ab initio} calculations reveal that, out of the $768$ non-equivalent vacancy structures, 164 systems exhibit DBSs pinned at the Fermi energy, while an additional 39 (32) 
{display vacancy levels located}
below (above) the Fermi level [Fig.~\ref{fig:painel}(d)]. 
{In the remaining structures, the vacancy levels are hybridized within} the non-defective electronic states, 
{precluding the formation of an isolated defect subspace.
This classification allows us to map} 
the vacancy-driven trajectories toward topological phase transitions 
as highlighted in the topological phase screening shown in Fig.~\ref{fig:painel}(e). 
{Depending on the host symmetry and the magnetic nature of the defect, the introduction of vacancies can either preserve or break time-reversal symmetry, dictating the nature of the resulting topological order.}
{Furthermore, the presence or absence of spatial inversion symmetry plays a critical role in the phase evolution, determining the sequence and multiplicity of transitions. 
We find that by maintaining either time-reversal or inversion symmetry, a Weyl semimetal phase frequently emerges as an intermediate regime prior to the realization of quantum anomalous Hall or quantum spin Hall states.}

%=======================  Figure 3 ==========================
%=======================  Figure 3 ==========================

\begin{figure*}[h!]
\includegraphics[width=\columnwidth, height=10.0cm]{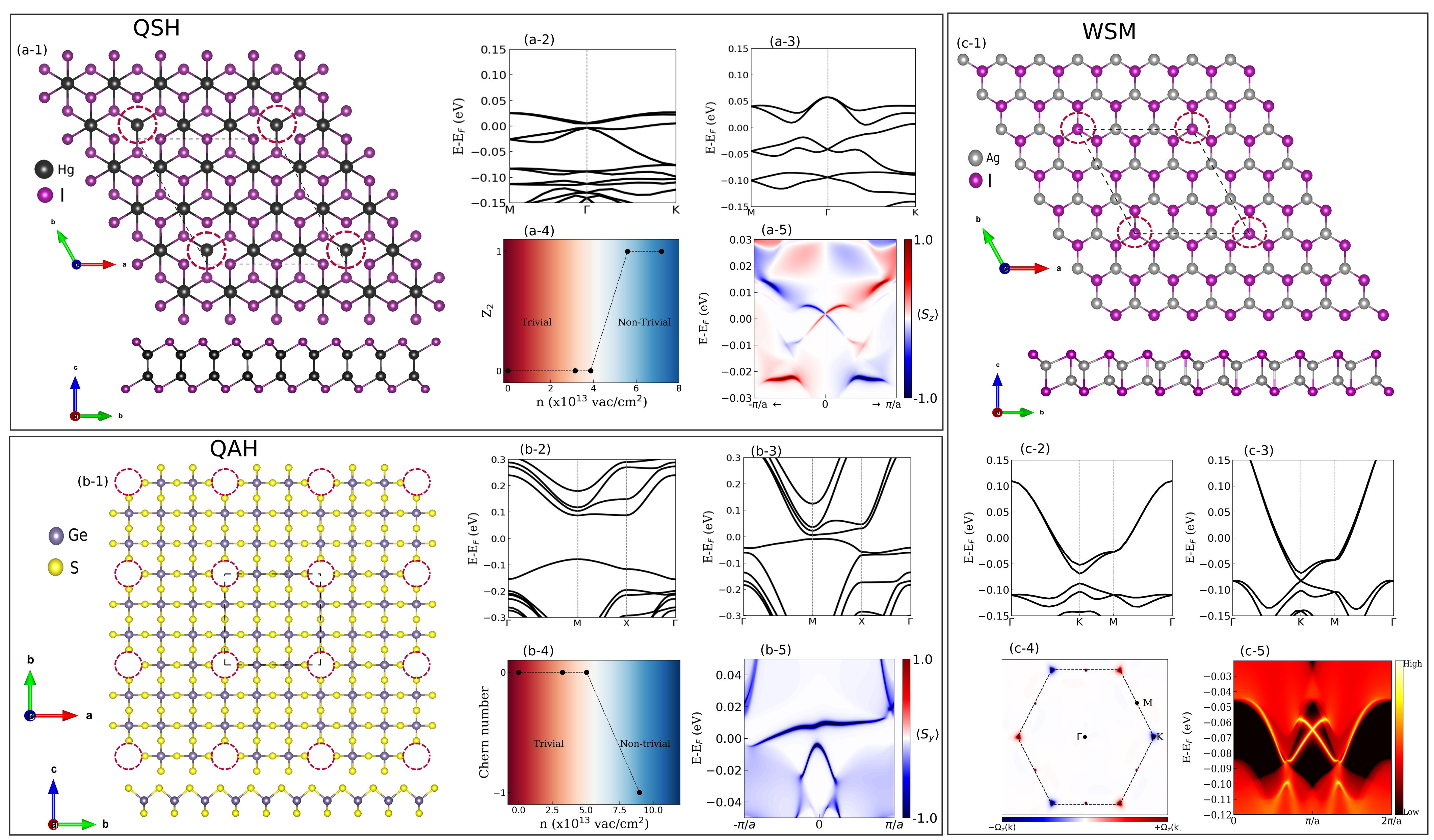}
\caption{\label{fig:example} {\bf Topological phase  driven by vacancy}. (a) Quantum spin Hall, (b) Quantum anomalous Hall, and (c) Weyl semimetal phase panels. Their respective chosen Vacancy structures are HgI (a-1), GeS$_2$ (b-1) and AgI (c-1) systems, with their corresponding band structures in low [(a-2), (b-2) and (c-2)] and medium [(a-3), (b-3) and (c-3)] vacancies concentration. The vacancy sites are indicated by red dashed circles. By increasing the vacancy concentration, the topological phase transition can be noted, as schematically shown in (a-4) and (b-4). The Weyl semimetal phase is a critical transition point, with a delta peak Berry curvature in Weyl points position (c-4). Due to bulk boundary-correspondence, spin-polarized topological edge states emerge connecting the bands with gap open by SOC in bulk (a-5) and (b-5). The topological edge states for WSM is shown in (c-5).}
\end{figure*}

As illustrated in Fig.~\ref{fig:painel}, 
{the targeted introduction of vacancies provides a robust pathway for inducing topological transitions in otherwise trivial semiconductors.
This process is governed by the competition between inter-vacancy hopping $t'$ and local spin-orbit strength $\lambda$.
To demonstrate this mechanism, we focus on}
the phase transition from trivial to QSH system and select a layered system composed by HgI [Fig. \ref{fig:example}(a)], which shows a trivial band gap at the order of 1.28 eV. 
By removing Hg atoms, DBS emerge {within the gap.} 
At low Hg-vacancy concentrations ($\eta = 3.14 \times 10^{13}$ vac/cm$^2$), the small magnitude of $t'$ do not
{drive a global reorganization of electronic} structure [Fig. \ref{fig:example}(a-2)].
The quantum spin Hall phase emerges in this system near $\eta = 5.6 \times 10^{13}$ vac/cm$^2$, where the interplay between of t' and spin-orbit strength leads to the non-trivial phase ($Z_{2}=1$) with usual inverted bands behavior [Fig. \ref{fig:example}(a-3)], when compared the band structures low Hg-vacancy counterpart. The topological phase transition as a function of vacancy concentration can be mapped in the phase diagram from [Fig. \ref{fig:example}(a-4)], where the non-trivial behavior persists at the concentration near to $10^{14}$ vac/cm$^2$. In the topological regime, the bulk-boundary correspondence can be visualized by means of a nanoribbon band structure [Fig. \ref{fig:example}(a-5)], where it can be noted spin-polarized edge states that cross each other at the Fermi level, which connecting the inverted bands in bulk. In the supplemental material, we present different systems, with different structures and atomic compositions, where the vacancy density increasing drives a topological phase transition.

Dangling bonds emerge in the defective systems due to vacancy introduction, which can lead to a non-vanishing magnetism, consequently breaking the time-reversal symmetry. In this way, the introduction of defects can also be a good mechanism to induce quantum anomalous Hall effect in different systems. As discussed in Methods, the introduction of the Zeeman term in effective Hamiltonian also suitable predict the topological phase transition for the magnetic regime under increasing vacancy density. Combined with the spin-orbit coupling and inter-vacancy hopping terms, the Zeeman  also rules the topological phase transition in different systems. In pristine form, GeS$_2$ monolayer is a trivial non-magnetic semiconductor with a sizable band gap ($1.57$ eV). Upon germanium vacancy introduction [Fig. \ref{fig:example}(b-1)], localized dangling-bond states emerge within the pristine gap and become partially spin-polarized due to unpaired electrons, effectively breaking time-reversal symmetry. By introducing the Ge-vacancy, a non-vanishing magnetism at the order of $0.47$~ $\mu_B$/S emerges in the structure. Under Ge-vacancy density of $\eta =8.98 \times 10^{13}$ vac/cm$^2$, the S-magnetic moment couples in the ferromagnetic (FM) phase for both nearest neighbor S ions ($\Delta{E}_{Intra}^{AFM-FM}= 20$ meV/S), and the inter-vacancy sites ($\Delta{E}_{Inter}^{AFM-FM} = 4$ meV/Vac). At lower densities, these DBSs remain weakly hybridized and the system preserves its trivial insulating character [Fig. \ref{fig:example}(b-2)].

As the vacancy density increases, the overlap between dangling-bond states becomes significant, leading to the formation of dispersive defect bands. In this regime, the combined effect of spin–orbit coupling and vacancy-induced magnetization opens a non-trivial gap characterized by a finite Chern number ($C=-1$), signaling the onset of a QAH phase [Fig. \ref{fig:example}(b-3)]. This transition is consistent with the effective tight-binding description, where the increase of the inter-vacancy hopping parameter $t'$ relative to the local spin–orbit coupling drives the system across a topological phase boundary. Figure \ref{fig:example}(b-4) illustrates the topological behavior as a function of Ge-vacancy density and Figure \ref{fig:example}(b-5) shows the corresponding spin-polarized edge state that emerges due to the bulk-boundary correspondence. Further increase in vacancy concentration enhances hybridization to the point where the defect bands merge and the gap closes, eventually reopening into a trivial phase, in agreement with the generic phase diagram predicted by the model.

 \begin{figure}[h!]
    \centering
    \includegraphics[width=0.8\linewidth]{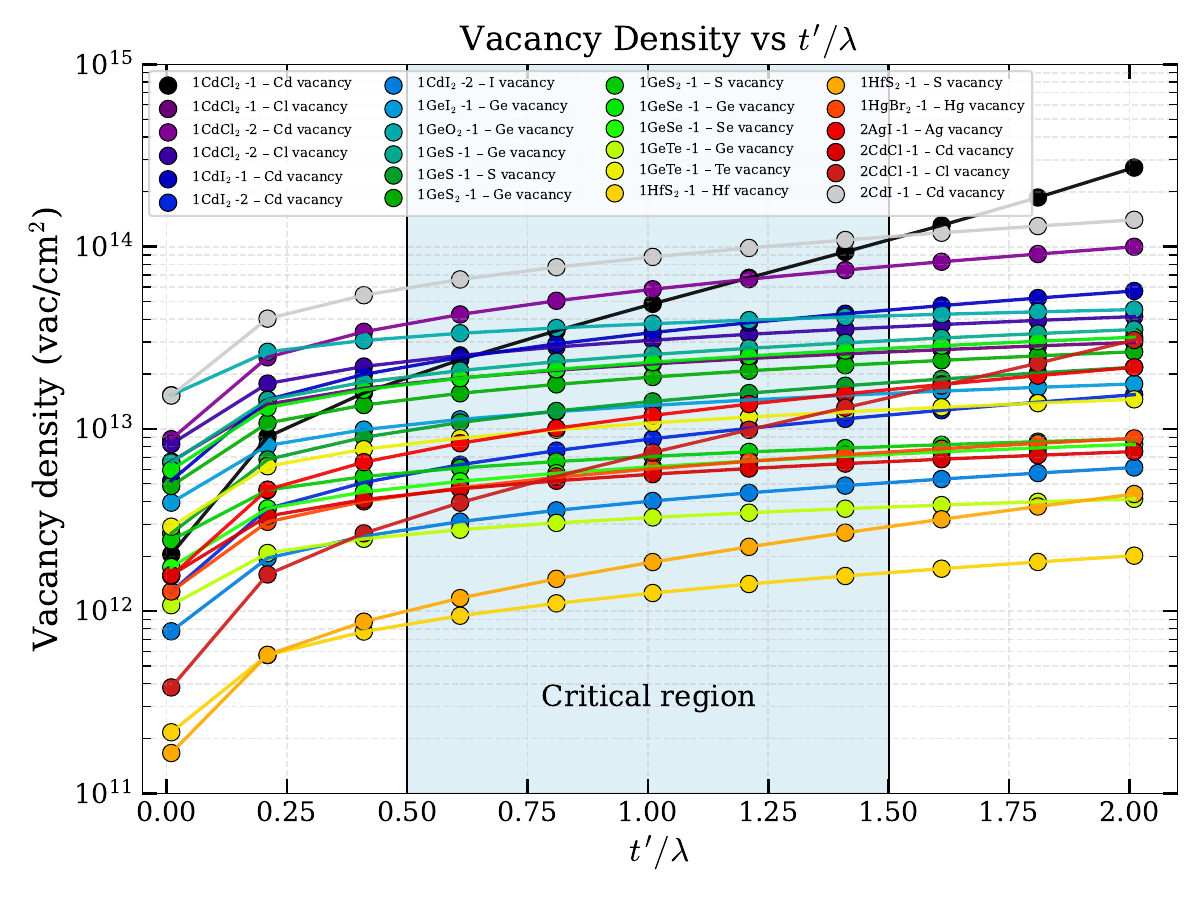}
    \caption{\label{fig:density} Critical vacancy density as a function of the $t'/\lambda$ value for different vacancy systems, as indicated in the legend box inset. Here, we employ the C2DB notation to name different structures with same atomic configuration. The critical region for the topological phase transition is marked in light blue. }
\end{figure}

The effective model predicts a critical point for the topological phase transition under increase in DBS interaction. Following the topological phase screening [Fig. \ref{fig:painel}(e)], a vacancy system that obeys a given symmetry route shows the 2D WSM phase in a given concentration. The possibility of 2D Weyl fermions due to DBS interaction has been confirmed by Cai \textit{et al.} in PtTe$_{1.75}$ system~\cite{NanolettCai2024}. In AgI system [Fig. \ref{fig:example}(c-1)], upon low Ag-vacancy concentration [Fig. \ref{fig:example}(c-2)], the system exhibits a trivial band gap located below the Fermi energy. At $\eta = 6.82\times 10^{13}$ vac/cm$^2$, the critical transition point takes place the gapped region [Fig. ~\ref{fig:example}(c-3)]. As DBS do not induce local magnetic moment and the inversion symmetry is broken, this crossing point is a good candidate for 2D WSM, as schematically illustrated in Fig. \ref{fig:painel}(e). By performing the 2D Weyl chirality ($\chi^{2D}$) calculations, we verify the crossing points non-trivial behavior with $\chi^{2D}=-1$ ($\chi^{2D}=+1$) at +K (-K) point. These results are further validated by the Berry curvature, which shows a delta peak centered at the position of topological crossing points [Fig. \ref{fig:example}(c-4)]. As can be seen in Fig. \ref{fig:example}(c-5), Topological edge states connecting Weyl points with opposite chiralities can be visualized by performing a nanoribbon calculation.

The same competition between local ordering and long-range coupling governs the behavior across the broader class of materials analyzed in our screening. As summarized in Fig.~\ref{fig:density}, the critical vacancy density required to induce a topological phase transition correlates with the effective inter-vacancy hopping extracted from first-principles calculations. Materials with more spatially extended dangling-bond orbitals require lower vacancy densities to reach the topological regime, while more localized defect states demand higher concentrations. Despite these quantitative differences, the qualitative mechanism remains unchanged: topology emerges from the collective reconstruction of vacancy-derived electronic states.

Taken together, these results demonstrate that vacancy-induced electronic reconstruction provides a universal route to engineer topological phases in two-dimensional semiconductors. Vacancies act neither as mere perturbations nor as sources of incoherent disorder, but rather as building blocks of an emergent electronic subspace whose topology is controlled by symmetry, spin–orbit coupling, and defect density. This framework naturally unifies defect physics and topological band theory, and establishes vacancy engineering as a realistic and experimentally accessible strategy to transform trivial materials into topological quantum matter.

\section*{Methods}\label{Methods}

\subsection*{Tight-Binding model}

Assuming $N$ identical localized sites, each defining a localized state $\{ \vert n \rangle \}$. Let those states form an orthonormal base $\langle n \vert m \rangle = \delta_{nm}$. The system Hamiltonian written in this base is 
\begin{equation}
    \hat{H}_0 = \begin{pmatrix}
        \langle 1 \vert \hat{H}_0 \vert 1 \rangle & \langle 1 \vert \hat{H}_0 \vert 2 \rangle & \cdots \\
        \langle 2 \vert \hat{H}_0 \vert 1 \rangle & \langle 2 \vert \hat{H}_0 \vert 2 \rangle & \cdots \\
        \vdots & \vdots & \ddots 
    \end{pmatrix}
\end{equation}

As the sites are identical $\langle n \vert \hat{H}_0 \vert n \rangle = \varepsilon_0$, which we will define as the energy reference $\varepsilon_0 = 0$. If two sites are adjacent (first neighbors) the hopping term will be 
\begin{equation}
    \langle n \vert \hat{H}_0 \vert n+1 \rangle = -t,
\end{equation}
and zero if not first neighbors.

\paragraph*{Spin-orbit coupling --} 
The electron (in the reference frame of the lab) can fell an electric field from the presence of a closer $j$'s site, as it hop from  the $n$ to the $n+1$ site. This electric field in the reference frame of the electron give rise to a magnetic field and couples with the spin degree of freedom (spin-orbit coupling). For a 2D system, where $n$, $n+1$ and $j$ sites are coplanar (i.e., in the $x-y$ plane) the effective magnetic field is along $z$ direction. Expanding our base to accommodate the spin degree of freedom $\{ \vert n \rangle \} \rightarrow \{ \vert n \uparrow \rangle, \vert n \downarrow \rangle \}$, the $\hat{H}_{soc}$ matrix representation has elements
\begin{equation}
    \langle n \uparrow \vert \hat{H}_{soc} \vert n+1\uparrow\rangle = i \nu_{n,n+1} \lambda,
\end{equation}
where $\nu_{n,n+1} = - \nu_{n+1,n}$ accommodate the direction dependence of the effective magnetic field sign. This $\hat{H}_{soc}$ for 2D system, giving its $\sigma_z$ proportionality, do not couples different spin subspaces.
For this reason we can analyze the Chern number of each spin subspace separately.
In the absence of time-reversal, with the spin subspaces energetically separated, the Chern number will define the topological invariant.
While for time reversal the spin-Chern number $C_s = C_{\uparrow} - C_{\downarrow}$ defines a topological invariant, where $C_{\sigma}$ is the Chern number computed in each spin-subspace.

\paragraph*{Zeeman term --}
If the system presents an spontaneous magnetization, or if is taken into account an external magnetic field (assuming to be perpendicular to the 2D plane), there will appear a Zeeman term
\begin{equation}
    \hat{H}_z = \mu {B}_{ext} \sigma_z.
\end{equation}
With the matrix elements in the previous defined base
\begin{equation}
    \langle n \uparrow \vert \hat{H}_z \vert n\uparrow \rangle = z ; \quad \langle n \downarrow \vert \hat{H}_z \vert n\downarrow \rangle = -z
\end{equation}
where $z$ is the energy scale.

\paragraph*{Periodic sites interaction --}
If this collection of sites are interacting with periodic images of it self there will also be a hopping term ($t'$). However given the periodicity an extra Bloch phase appears into the original cell Hamiltonian,
\begin{equation}
    \langle n \vert \hat{H}_p \vert m \rangle = t' e^{i\vec{k} \cdot \vec{R}_{m,n}},
\end{equation}
where $\vec{R}$ is the vector connecting the periodic cells.

\paragraph*{Full Hamiltonian analysis --}
Here the periodic site interaction is geometry dependent. Let the real space lattice, described by $\vec{R} = r\vec{a}_1 + s \vec{a}_2 $. The predominant inter vacancy interaction will be between adjacent cells, that is, $r,s = 0,\pm1$.  Therefore writing $\vec{k} = x\vec{b}_1 + y \vec{b}_2$, with $x,y \in [0,1)$ in the first Brillouin zone, we have the possible values the Bloch phase $f_{mn}(\vec{k}) = e^{i\vec{k} \cdot \vec{R}} \in \left\{ e^{\pm i 2\pi x},e^{\pm i 2\pi y}, e^{\pm i 2\pi (x+y)}, e^{\pm i 2\pi (x-y)}  \right\}$. In order to see if a geometry dependence arises, for each vacancy local environment (ring with N sites), we solve the Hamiltonian for 1000 different systems by randomly selecting one of the possible Bloch phase for each inter-site coupling. That is for random choice of $f_{mn}(\vec{k})$ in the Hamiltonian
\begin{equation}
\hat{H}_N = \begin{pmatrix} 0 & t+i\lambda & 0 & \cdots \\ t-i\lambda & 0 & t+i\lambda & \cdots \\ 0 & t-i\lambda & 0 & \cdots \\ \vdots & \vdots & \vdots & \ddots \end{pmatrix} + t' \begin{pmatrix} f_{11} & f_{12} & \cdots \\ f_{21} & f_{22} & \cdots \\ \vdots & \vdots & \ddots  \end{pmatrix}.
\end{equation}
We can solve the Hamiltonian and for each k point we can calculate the accumulated phase over the unity cell ring. This phase can be obtained by the expected values of an operator that define the cycle $n \rightarrow n+1$,
\begin{equation}
    \hat{P} = \begin{pmatrix}
        0 & 1 & 0  & \cdots \\
        0 & 0 & 1 & \cdots \\
        \vdots & \vdots & \ddots & \\
        1 & 0 & \cdots & 0
    \end{pmatrix} = \sum_n^{N} \vert n+1 \rangle \langle n \vert
\end{equation}
Here, $\vert N+1\rangle \equiv \vert 1\rangle$. The accumulated phase is
\begin{equation}
    \frac{1}{2\pi}\arg \left( \langle \psi \vert P \vert \psi \rangle \right).
\end{equation}

In order to validate this gap inversion, we have computed the spin Chern number within this gap for the generated 1000 system. 

\subsection*{Effective analysis}

We can understand the behavior observed by the random periodic 1000 system by a simpler model. Taking all those term together we see that they define two separated/orthogonal subspaces (spin-up and spin-down), with the matrix elements for adjacent sites being with the form
\begin{eqnarray}
    \Delta &=& t\pm i \lambda  + t' e^{i\vec{k}\cdot \vec{R}} \\
    &=& [t + t' \cos (\vec{k} \cdot \vec{R})] + i [\pm \lambda +t' \sin (\vec{k} \cdot \vec{R})]
\end{eqnarray}
That is containing real and imaginary parts. The values of the real and imaginary parts depend on the system couplings.
Writing $\Delta = e^{i\phi}$, in the complex plane representation each term is a vector shown in Fig.~\ref{fig:vacancy-phase}(a).

Lets analyze this system from a configuration of N-site ring, each coupled with its adjacent site. The general Hamiltonian (for spin-up subspace) will be
\begin{equation}
    \hat{H}_{N\uparrow} = \begin{pmatrix}
        0 & \Delta & 0 & \cdots & \Delta^* \\
        \Delta^* & 0 & \Delta & \cdots & 0 \\
        0 & \Delta^* & 0 & \cdots & 0 \\
        \vdots & \vdots & \vdots & \ddots & \vdots \\
        \Delta &0 & 0 & \cdots & 0
    \end{pmatrix}.
\end{equation}
This is a circulant matrix (each line is the right shift if the line above) and has eigenvectors as the Fourier modes
\begin{equation}
    \vert j \rangle = \frac{1}{\sqrt{N}} \begin{pmatrix}
        1 \\ w^{j} \\ w^{2j} \\ \vdots \\ w^{(N-1)j}
    \end{pmatrix}; \quad w=e^{i\frac{2\pi}{N}}
\end{equation}
and eigenvalues 
\begin{equation}
    \lambda_j = 2 \cos \left( \phi + \frac{2\pi}{N} j  \right),
\end{equation}
with $j=0,1,2,\cdots,(N-1)$. 

For $\phi =0$, that is $\Delta  \in \mathbb{R}$, we have for $N$ even, two non-degenerates states ($j=0$ and $j=N/2$), and $N/2 -1$ doubledegenerated states pairing $j=n$ with $j=N-n$. For $N$ odd the same doubledegenerated pairing is presented but existing only one non-degenerated state for $j=0$.

We can define for each of those states an accumulated phase over a cycle on the system ring. This phase can be obtained by the expected values of an operator that define the cycle $n \rightarrow n+1$,
\begin{equation}
    \hat{P} = \begin{pmatrix}
        0 & 1 & 0  & \cdots \\
        0 & 0 & 1 & \cdots \\
        \vdots & \vdots & \ddots & \\
        1 & 0 & \cdots & 0
    \end{pmatrix} = \sum_n \vert n+1 \rangle \langle n \vert
\end{equation}
Here, $\vert N+1\rangle \equiv \vert 1\rangle$. This expectation value is identical to the sum of the right diagonal of the density matrix $\vert j \rangle \langle j \vert$,
\begin{equation}
    \langle j \vert \hat{P} \vert j \rangle = (w^j)^* = e^{-i 2 \pi j /N}.
\end{equation}
Notice that a given degenerated pair $[n,N-n]$, have accumulated phase with opposing signal $(w^{N-n})^* = w^n$.

For $\phi \neq 0$, that is $\Delta$ has an imaginary part, the phase accumulated by each states are still the same.  However, there is a breaking of the doubledegenerated pairs. The energy ordering of those pair of states depends on the signal of $\phi$. There is not a way of inverting this order without passing through the doubledegenerated state. In other words, there is not an adiabatic evolution connecting the two gap regimes without closing the gap. Therefore those two regimes have different topologies. 

Lets look closer to the phase accumulation for each state, that is the expectation value o $\hat{P}$ operator. First defining a local current operator
\begin{equation}
    \hat{J}_{ij} = \frac{i}{\hbar} \left( t_{ij} \vert i \rangle \langle j \vert - t_{ji} \vert j \rangle \langle i \vert \right)
\end{equation}
where $t_{ij} = t_{ji} = t$ for $i$ adjacent to $j$. The total current on the ring is 
\begin{eqnarray*}
    \hat{J} = \sum_n J_{n.n+1} &=& \frac{i t}{\hbar} \sum_n \left( \vert n \rangle \langle n+1 \vert - \vert n+1 \rangle \langle n \vert \right) \\
    &=& \frac{it}{\hbar} (P^T - P)
\end{eqnarray*}
Calculating the expectation value, where are $\hat{P}$ is real $\langle j \vert \hat{P}^T \vert j \rangle = \langle j \vert \hat{P} \vert j \rangle  ^*$, therefore
\begin{equation}
    \langle \hat{J} \rangle = \frac{t}{\hbar} 2 {\rm Im} \left( \langle j \vert \hat{P} \vert j \rangle  \right).
\end{equation}
That is, the imaginary part of the $\hat{P}$ operator is related to the current circulating the ring, which is defined by the accumulated phase. Here spin-up and spin-down have opposing current direction.

\subsection*{Density Functional Theory simulations}

All systems were simulated within the Vienna {\it ab initio} simulation package (VASP) \cite{PRBkresse1996}, using density functional theory (DFT) in the generalized gradient approximation of Perdew-Burke-Ernzerhof (PBE) \cite{PRLperdew1996}. The plane-wave basis was truncated at a kinetic energy cutoff of $400$\,eV, and Brillouin-zone integrations were performed on Monkhorst-Pack meshes \cite{PRBmonkhorst1976} of $2/${\AA}$^{-1}$ density in each reciprocal direction. The electron–ion interaction was treated with the projector augmented-wave (PAW) method \cite{PRBblochl1994}, while ionic positions were relaxed until the residual forces were smaller than $0.5 \times 10^{-3}$\,eV/{\AA} and the total electronic energy converged below $10^{-6}$\,eV. The topological properties were calculated in the Wannier functions framework, extracted from DFT within the Wannier90 package \cite{Wannier90}.

\section*{Data availability}

\section*{Acknowledgements}
The authors acknowledge financial support from the Brazilian agencies FAPESP (23/09820-2), CNPq (INCT - Materials Informatics), and LNCC (Laboratório Nacional de Computação Científica) for computer time (Project DIDMat).

\bibliography{refs}% Produces the bibliography via BibTeX.

@article{NATNANOfang2023,
  title = {{Atomically precise vacancy-assembled quantum antidots}},
  volume = {18},
  ISSN = {1748-3395},
  url = {http://dx.doi.org/10.1038/s41565-023-01495-z},
  DOI = {10.1038/s41565-023-01495-z},
  number = {12},
  journal = {Nat. Nanotechnol.},
  publisher = {Springer Science and Business Media LLC},
  author = {Fang,  Hanyan and Mahalingam,  Harshitra and Li,  Xinzhe and Han,  Xu and Qiu,  Zhizhan and Han,  Yixuan and Noori,  Keian and Dulal,  Dikshant and Chen,  Hongfei and Lyu,  Pin and Yang,  Tianhao and Li,  Jing and Su,  Chenliang and Chen,  Wei and Cai,  Yongqing and Neto,  A. H. Castro and Novoselov,  Kostya S. and Rodin,  Aleksandr and Lu,  Jiong},
  year = {2023},
  month = aug,
  pages = {1401–1408}
}

@article{PRBkohan2000,
  title = {{First-principles study of native point defects in ZnO}},
  author = {Kohan, A. F. and Ceder, G. and Morgan, D. and Van de Walle, Chris G.},
  journal = {Phys. Rev. B},
  volume = {61},
  issue = {22},
  pages = {15019--15027},
  numpages = {0},
  year = {2000},
  month = {Jun},
  publisher = {American Physical Society},
  doi = {10.1103/PhysRevB.61.15019},
  url = {https://link.aps.org/doi/10.1103/PhysRevB.61.15019}
}

@article{PRLzhang1991,
  title = {{Chemical potential dependence of defect formation energies in GaAs: Application to Ga self-diffusion}},
  author = {Zhang, S. B. and Northrup, John E.},
  journal = {Phys. Rev. Lett.},
  volume = {67},
  issue = {17},
  pages = {2339--2342},
  numpages = {0},
  year = {1991},
  month = {Oct},
  publisher = {American Physical Society},
  doi = {10.1103/PhysRevLett.67.2339},
  url = {https://link.aps.org/doi/10.1103/PhysRevLett.67.2339}
}

@article{ACSNANOnan2014,
  title = {Strong Photoluminescence Enhancement of {MoS}$_2$ through Defect Engineering and Oxygen Bonding},
  volume = {8},
  ISSN = {1936-086X},
  url = {http://dx.doi.org/10.1021/nn500532f},
  DOI = {10.1021/nn500532f},
  number = {6},
  journal = {ACS Nano},
  publisher = {American Chemical Society (ACS)},
  author = {Nan, H. and Wang, Z. and Wang, W. and Liang, Z. and Lu, Y. and Chen, Q. and He, D. and Tan, P. and Miao, F. and Wang, X. and Wang, J. and Ni, Z.},
  year = {2014},
  month = may,
  pages = {5738–5745}
}

@article{NATCOMMqiu2013,
  title = {{Hopping transport through defect-induced localized states in molybdenum disulphide}},
  volume = {4},
  pages ={2642},
  ISSN = {2041-1723},
  url = {http://dx.doi.org/10.1038/ncomms3642},
  DOI = {10.1038/ncomms3642},
  number = {1},
  journal = {Nat. Commun.},
  publisher = {Springer Science and Business Media LLC},
  author = {Qiu, H. and Xu, T. and Wang, Z. and Ren, W. and Nan, H. and Ni, Z. and Chen, Q. and Yuan,  S. and Miao, F. and Song, F. and Long, G. and Shi, Y. and Sun, L. and Wang, J. and Wang, X.},
  year = {2013},
  month = oct 
}

@article{PRBmosca2023,
  title = {{High-temperature vacancy-induced magnetism in nanostructured materials}},
  author = {Mosca, D. H. and Varalda, J. and Dartora, C. A.},
  journal = {Phys. Rev. B},
  volume = {108},
  issue = {13},
  pages = {134410},
  numpages = {12},
  year = {2023},
  month = {Oct},
  publisher = {American Physical Society},
  doi = {10.1103/PhysRevB.108.134410},
  url = {https://link.aps.org/doi/10.1103/PhysRevB.108.134410}
}

@article{PRLkomsa2012,
  title = {Two-Dimensional Transition Metal Dichalcogenides under Electron Irradiation: Defect Production and Doping},
  author = {Komsa, H.-P. and Kotakoski, J. and Kurasch, S. and Lehtinen, O. and Kaiser, U. and Krasheninnikov, A. V.},
  journal = {Phys. Rev. Lett.},
  volume = {109},
  issue = {3},
  pages = {035503},
  numpages = {5},
  year = {2012},
  month = {Jul},
  publisher = {American Physical Society},
  doi = {10.1103/PhysRevLett.109.035503},
  url = {https://link.aps.org/doi/10.1103/PhysRevLett.109.035503}
}

@article{JPDAParaujo2022,
  title = {{Unveiling ferromagnetism and antiferromagnetism in two dimensions at room temperature}},
  volume = {55},
  ISSN = {1361-6463},
  url = {http://dx.doi.org/10.1088/1361-6463/ac60cd},
  DOI = {10.1088/1361-6463/ac60cd},
  number = {28},
  journal = {J. Phys. D: Appl. Phys.},
  publisher = {IOP Publishing},
  author = {Araujo, R. de Moraes Telles and Zarpellon, J. and Mosca, D. H.},
  year = {2022},
  month = apr,
  pages = {283003}
}

@article{RMPhasan2010,
  title = {{Colloquium: Topological insulators}},
  author = {Hasan, M. Z. and Kane, C. L.},
  journal = {Rev. Mod. Phys.},
  volume = {82},
  issue = {4},
  pages = {3045--3067},
  numpages = {0},
  year = {2010},
  month = {Nov},
  publisher = {American Physical Society},
  doi = {10.1103/RevModPhys.82.3045},
  url = {https://link.aps.org/doi/10.1103/RevModPhys.82.3045}
}

@article{RMPQi2010,
  title = {Topological insulators and superconductors},
  author = {Qi, X.-L. and Zhang, S.-C.},
  journal = {Rev. Mod. Phys.},
  volume = {83},
  issue = {4},
  pages = {1057--1110},
  numpages = {0},
  year = {2011},
  month = {Oct},
  publisher = {American Physical Society},
  doi = {10.1103/RevModPhys.83.1057},
  url = {https://link.aps.org/doi/10.1103/RevModPhys.83.1057}
}

@misc{c2db,
  doi = {10.11583/DTU.14616660.V1},
  url = {https://data.dtu.dk/articles/dataset/Computational_2D_Materials_Database_C2DB_/14616660/1},
  author = {Thygesen,  Kristian Sommer},
  title = {{Computational 2D Materials Database (C2DB)}},
  publisher = {Technical University of Denmark},
  year = {2021},
  copyright = {Creative Commons Attribution 4.0 International}
}

@article{PRBkresse1996,
  title = {{Efficient iterative schemes for ab initio total-energy calculations using a plane-wave basis set}},
  author = {Kresse, G. and Furthm\"uller, J.},
  journal = {Phys. Rev. B},
  volume = {54},
  issue = {16},
  pages = {11169--11186},
  numpages = {0},
  year = {1996},
  month = {Oct},
  publisher = {American Physical Society},
  doi = {10.1103/PhysRevB.54.11169},
  url = {https://link.aps.org/doi/10.1103/PhysRevB.54.11169}
}

@article{PRLperdew1996,
  title = {{Generalized Gradient Approximation Made Simple}},
  author = {Perdew, J. P. and Burke, K. and Ernzerhof, M.},
  journal = {Phys. Rev. Lett.},
  volume = {77},
  issue = {18},
  pages = {3865--3868},
  numpages = {0},
  year = {1996},
  month = {Oct},
  publisher = {American Physical Society},
  doi = {10.1103/PhysRevLett.77.3865},
  url = {https://link.aps.org/doi/10.1103/PhysRevLett.77.3865}
}

@article{PRBmonkhorst1976,
  title = {{Special points for Brillouin-zone integrations}},
  author = {Monkhorst, H. J. and Pack, J. D.},
  journal = {Phys. Rev. B},
  volume = {13},
  issue = {12},
  pages = {5188--5192},
  numpages = {0},
  year = {1976},
  month = {Jun},
  publisher = {American Physical Society},
  doi = {10.1103/PhysRevB.13.5188},
  url = {https://link.aps.org/doi/10.1103/PhysRevB.13.5188}
}

@article{PRLkane2005,
  title = {{Quantum Spin Hall Effect in Graphene}},
  author = {Kane, C. L. and Mele, E. J.},
  journal = {Phys. Rev. Lett.},
  volume = {95},
  issue = {22},
  pages = {226801},
  numpages = {4},
  year = {2005},
  month = {Nov},
  publisher = {American Physical Society},
  doi = {10.1103/PhysRevLett.95.226801},
  url = {https://link.aps.org/doi/10.1103/PhysRevLett.95.226801}
}

@article{PRLli2009,
  title = {{Topological Anderson Insulator}},
  author = {Li, J. and Chu, R.-L. and Jain, J. K. and Shen, S.-Q.},
  journal = {Phys. Rev. Lett.},
  volume = {102},
  issue = {13},
  pages = {136806},
  numpages = {4},
  year = {2009},
  month = {Apr},
  publisher = {American Physical Society},
  doi = {10.1103/PhysRevLett.102.136806},
  url = {https://link.aps.org/doi/10.1103/PhysRevLett.102.136806}
}

@article{PRLgroth2009,
  title = {{Theory of the Topological Anderson Insulator}},
  author = {Groth, C. W. and Wimmer, M. and Akhmerov, A. R. and Tworzyd\l{}o, J. and Beenakker, C. W. J.},
  journal = {Phys. Rev. Lett.},
  volume = {103},
  issue = {19},
  pages = {196805},
  numpages = {4},
  year = {2009},
  month = {Nov},
  publisher = {American Physical Society},
  doi = {10.1103/PhysRevLett.103.196805},
  url = {https://link.aps.org/doi/10.1103/PhysRevLett.103.196805}
}

@article{PRBL-Regis2024,
  title = {Structure-driven phase transitions in paracrystalline topological insulators},
  author = {Regis, V. and Velasco, V. and {Silva Neto}, M. B. and Lewenkopf, C.},
  journal = {Phys. Rev. B},
  volume = {110},
  issue = {16},
  pages = {L161105},
  numpages = {6},
  year = {2024},
  month = {Oct},
  publisher = {American Physical Society},
  doi = {10.1103/PhysRevB.110.L161105},
  url = {https://link.aps.org/doi/10.1103/PhysRevB.110.L161105}
}

@article{PRBL-Assuncao2024,
  title = {Phase transitions and scale invariance in topological {Anderson} insulators},
  author = {Assun\c{c}\~ao, B. D. and Ferreira, G. J. and Lewenkopf, C. H.},
  journal = {Phys. Rev. B},
  volume = {109},
  issue = {20},
  pages = {L201102},
  numpages = {7},
  year = {2024},
  month = {May},
  publisher = {American Physical Society},
  doi = {10.1103/PhysRevB.109.L201102},
  url = {https://link.aps.org/doi/10.1103/PhysRevB.109.L201102}
}

@article{PRBfocassio2021,
  title = {{Amorphous ${\mathrm{Bi}}_{2}{\mathrm{Se}}_{3}$ structural, electronic, and topological nature from first principles}},
  author = {Focassio, Bruno and Schleder, Gabriel R. and Crasto de Lima, Felipe and Lewenkopf, Caio and Fazzio, Adalberto},
  journal = {Phys. Rev. B},
  volume = {104},
  issue = {21},
  pages = {214206},
  numpages = {11},
  year = {2021},
  month = {Dec},
  publisher = {American Physical Society},
  doi = {10.1103/PhysRevB.104.214206},
  url = {https://link.aps.org/doi/10.1103/PhysRevB.104.214206}
}

@article{PRLagarwala2017,
  title = {{Topological Insulators in Amorphous Systems}},
  author = {Agarwala, Adhip and Shenoy, Vijay B.},
  journal = {Phys. Rev. Lett.},
  volume = {118},
  issue = {23},
  pages = {236402},
  numpages = {6},
  year = {2017},
  month = {Jun},
  publisher = {American Physical Society},
  doi = {10.1103/PhysRevLett.118.236402},
  url = {https://link.aps.org/doi/10.1103/PhysRevLett.118.236402}
}

@article{PRBblochl1994,
  title = {{Projector augmented-wave method}},
  author = {Bl\"ochl, P. E.},
  journal = {Phys. Rev. B},
  volume = {50},
  issue = {24},
  pages = {17953--17979},
  numpages = {0},
  year = {1994},
  month = {Dec},
  publisher = {American Physical Society},
  doi = {10.1103/PhysRevB.50.17953},
  url = {https://link.aps.org/doi/10.1103/PhysRevB.50.17953}
}

@article{NATNANOyang2008,
  title = {{Memristive switching mechanism for metal/oxide/metal nanodevices}},
  volume = {3},
  ISSN = {1748-3395},
  url = {http://dx.doi.org/10.1038/nnano.2008.160},
  DOI = {10.1038/nnano.2008.160},
  number = {7},
  journal = {Nat. Nanotechnol.},
  publisher = {Springer Science and Business Media LLC},
  author = {Yang,  J. Joshua and Pickett,  Matthew D. and Li,  Xuema and Ohlberg,  D. A. A. and Stewart,  D. R. and Williams,  R. Stanley},
  year = {2008},
  month = jun,
  pages = {429–433}
}

@article{PRshockley1952,
  title = {{Statistics of the Recombinations of Holes and Electrons}},
  author = {Shockley, W. and Read, W. T.},
  journal = {Phys. Rev.},
  volume = {87},
  issue = {5},
  pages = {835--842},
  numpages = {0},
  year = {1952},
  month = {Sep},
  publisher = {American Physical Society},
  doi = {10.1103/PhysRev.87.835},
  url = {https://link.aps.org/doi/10.1103/PhysRev.87.835}
}

@article{SCIENCEbernevig2006,
  title = {{Quantum Spin Hall Effect and Topological Phase Transition in HgTe Quantum Wells}},
  volume = {314},
  ISSN = {1095-9203},
  url = {http://dx.doi.org/10.1126/science.1133734},
  DOI = {10.1126/science.1133734},
  number = {5806},
  journal = {Science},
  publisher = {American Association for the Advancement of Science (AAAS)},
  author = {Bernevig,  B. Andrei and Hughes,  Taylor L. and Zhang,  Shou-Cheng},
  year = {2006},
  month = dec,
  pages = {1757–1761}
}

@article{AMTneves2025,
author = {das Neves, M. F. F. and Barêa, H. M. and Perfecto, T. and Bettini, J. and de Lima, F. C. and Oliveira, R. F. and Fazzio, A. and Leite, E. R. and Santhiago, M.},
title = {{Room-Temperature Tuning of Electrical Conductivity in Single MoS$_2$ Flakes via Nanoscale Amorphization by Focused Ion Beam}},
journal = {Advanced Materials Technologies},
volume = {10},
number = {20},
pages = {e01505},
doi = {https://doi.org/10.1002/admt.202501505},
year = {2025}
}

@article{NATMATlin2017,
  title = {{Intrinsically patterned two-dimensional materials for selective adsorption of molecules and nanoclusters}},
  volume = {16},
  ISSN = {1476-4660},
  url = {http://dx.doi.org/10.1038/nmat4915},
  DOI = {10.1038/nmat4915},
  number = {7},
  journal = {Nat. Mater.},
  publisher = {Springer Science and Business Media LLC},
  author = {Lin,  X. and Lu,  J. C. and Shao,  Y. and Zhang,  Y. Y. and Wu,  X. and Pan,  J. B. and Gao,  L. and Zhu,  S. Y. and Qian,  K. and Zhang,  Y. F. and Bao,  D. L. and Li,  L. F. and Wang,  Y. Q. and Liu,  Z. L. and Sun,  J. T. and Lei,  T. and Liu,  C. and Wang,  J. O. and Ibrahim,  K. and Leonard,  D. N. and Zhou,  W. and Guo,  H. M. and Wang,  Y. L. and Du,  S. X. and Pantelides,  S. T. and Gao,  H.-J.},
  year = {2017},
  month = jun,
  pages = {717–721}
}

@article{NATCOMMwang2017,
  title = {{Room temperature magnetization switching in topological insulator-ferromagnet heterostructures by spin-orbit torques}},
  volume = {8},
  url = {http://dx.doi.org/10.1038/s41467-017-01583-4},
  doi = {10.1038/s41467-017-01583-4},
  number = {1},
  journal = {Nat. Commun.},
  publisher = {Springer Science and Business Media LLC},
  author = {Wang,  Yi and Zhu,  Dapeng and Wu,  Yang and Yang,  Yumeng and Yu,  Jiawei and Ramaswamy,  Rajagopalan and Mishra,  Rahul and Shi,  Shuyuan and Elyasi,  Mehrdad and Teo,  Kie-Leong and Wu,  Yihong and Yang,  Hyunsoo},
  year = {2017},
  pages = {1364}
}

@article{SSCedelstein1990,
  title = {{Spin polarization of conduction electrons induced by electric current in two-dimensional asymmetric electron systems}},
  volume = {73},
  ISSN = {0038-1098},
  url = {http://dx.doi.org/10.1016/0038-1098(90)90963-C},
  doi = {10.1016/0038-1098(90)90963-c},
  number = {3},
  journal = {Solid State Commun.},
  publisher = {Elsevier BV},
  author = {Edelstein,  V. M.},
  year = {1990},
  month = jan,
  pages = {233–235}
}

@article{NATMATfan2014,
  title = {{Magnetization switching through giant spin–orbit torque in a magnetically doped topological insulator heterostructure}},
  volume = {13},
  ISSN = {1476-4660},
  url = {http://dx.doi.org/10.1038/nmat3973},
  DOI = {10.1038/nmat3973},
  number = {7},
  journal = {Nat. Mater.},
  publisher = {Springer Science and Business Media LLC},
  author = {Fan, Y. and Upadhyaya, P. and Kou, X. and Lang,  M. and Takei, S. and Wang,  Z. and Tang, J. and He,  L. and Chang,  L.-T. and Montazeri, M. and Yu, G. and Jiang, W. and Nie, T. and Schwartz, R. N. and Tserkovnyak, Y. and Wang, K. L.},
  year = {2014},
  month = apr,
  pages = {699–704}
}

@article{NATmanipatruni2018,
  title = {{Scalable energy-efficient magnetoelectric spin–orbit logic}},
  volume = {565},
  ISSN = {1476-4687},
  url = {http://dx.doi.org/10.1038/s41586-018-0770-2},
  DOI = {10.1038/s41586-018-0770-2},
  number = {7737},
  journal = {Nature},
  publisher = {Springer Science and Business Media LLC},
  author = {Manipatruni, S. and Nikonov, D. E. and Lin, C.-C. and Gosavi, T. A. and Liu, H. and Prasad,  B. and Huang, Y.-L. and Bonturim, E. and Ramesh, R. and Young, I. A.},
  year = {2018},
  month = dec,
  pages = {35–42}
}

@article{PhysRevB.96.115407,
  title = {Long-range entanglement for spin qubits via quantum Hall edge modes},
  author = {Elman, Samuel J. and Bartlett, Stephen D. and Doherty, Andrew C.},
  journal = {Phys. Rev. B},
  volume = {96},
  issue = {11},
  pages = {115407},
  numpages = {12},
  year = {2017},
  month = {Sep},
  publisher = {American Physical Society},
  doi = {10.1103/PhysRevB.96.115407},
  url = {https://link.aps.org/doi/10.1103/PhysRevB.96.115407}
}

@article{Ohfuchi2023,
  title = {Quantum Spin Hall States in 2D Monolayer WTe2/MoTe2 Lateral Heterojunctions for Topological Quantum Computation},
  volume = {6},
  ISSN = {2574-0970},
  url = {http://dx.doi.org/10.1021/acsanm.2c05027},
  DOI = {10.1021/acsanm.2c05027},
  number = {3},
  journal = {ACS Applied Nano Materials},
  publisher = {American Chemical Society (ACS)},
  author = {Ohfuchi,  Mari and Sekine,  Akihiko},
  year = {2023},
  month = jan,
  pages = {2020–2026}
}

@article{Xiao2023,
  title = {Utilizing Topological Insulator Two‐Dimensional Bismuth for Ultrasensitive Acoustic Detection},
  volume = {19},
  ISSN = {1613-6829},
  url = {http://dx.doi.org/10.1002/smll.202303608},
  DOI = {10.1002/smll.202303608},
  number = {49},
  journal = {Small},
  publisher = {Wiley},
  author = {Xiao,  Qi and Ma,  Bo and Wang,  Shu‐Yan and Li,  Xiang‐Yang and Yan,  Feng and Wang,  Qiang and Zhang,  Hao‐Li},
  year = {2023},
  month = aug 
}

@article{Li2022,
  title = {Filling the gap between topological insulator nanomaterials and triboelectric nanogenerators},
  volume = {13},
  ISSN = {2041-1723},
  url = {http://dx.doi.org/10.1038/s41467-022-28575-3},
  DOI = {10.1038/s41467-022-28575-3},
  number = {1},
  journal = {Nature Communications},
  publisher = {Springer Science and Business Media LLC},
  author = {Li,  Mengjiao and Lu,  Hong-Wei and Wang,  Shu-Wei and Li,  Rei-Ping and Chen,  Jiann-Yeu and Chuang,  Wen-Shuo and Yang,  Feng-Shou and Lin,  Yen-Fu and Chen,  Chih-Yen and Lai,  Ying-Chih},
  year = {2022},
  month = feb 
}

@article{Yan2024,
  title = {{Rules for dissipationless topotronics}},
  volume = {10},
  ISSN = {2375-2548},
  url = {http://dx.doi.org/10.1126/sciadv.ado4756},
  DOI = {10.1126/sciadv.ado4756},
  number = {23},
  journal = {Science Advances},
  publisher = {American Association for the Advancement of Science (AAAS)},
  author = {Yan,  Qing and Li,  Hailong and Jiang,  Hua and Sun,  Qing-Feng and Xie,  X. C.},
  year = {2024},
  month = jun 
}

@article{NATPHYSxia2009,
  title = {{Observation of a large-gap topological-insulator class with a single Dirac cone on the surface}},
  volume = {5},
  ISSN = {1745-2481},
  url = {http://dx.doi.org/10.1038/nphys1274},
  DOI = {10.1038/nphys1274},
  number = {6},
  journal = {Nat. Phys.},
  publisher = {Springer Science and Business Media LLC},
  author = {Xia, Y. and Qian, D. and Hsieh, D. and Wray, L. and Pal, A. and Lin, H. and Bansil, A. and Grauer, D. and Hor, Y. S. and Cava, R. J. and Hasan, M. Z.},
  year = {2009},
  month = may,
  pages = {398–402}
}

@article{NPcaio2019,
  title = {{Topological marker currents in Chern insulators}},
  volume = {15},
  ISSN = {1745-2481},
  url = {http://dx.doi.org/10.1038/s41567-018-0390-7},
  DOI = {10.1038/s41567-018-0390-7},
  number = {3},
  journal = {Nat. Phys.},
  publisher = {Springer Science and Business Media LLC},
  author = {Caio,  M. D. and M\"{o}ller,  G. and Cooper,  N. R. and Bhaseen,  M. J.},
  year = {2019},
  month = jan,
  pages = {257–261}
}

@article{NanolettCai2024,
  title = {Evidence for Two-Dimensional {Weyl} Fermions in Air-Stable Monolayer {PtTe}$_{1.75}$},
  volume = {24},
  ISSN = {1530-6992},
  url = {http://dx.doi.org/10.1021/acs.nanolett.4c02580},
  DOI = {10.1021/acs.nanolett.4c02580},
  number = {33},
  journal = {Nano Lett.},
  publisher = {American Chemical Society (ACS)},
  author = {Cai, Z. and Cao, H. and Sheng, H. and Hu, X. and Sun, Z. and Zhao, Q. and Gao, J. and Ideta,  S.-i. and Shimada, K. and Huang, J. and Cheng, P. and Chen, L. and Yao, Y. and Meng, S. and Wu, K. and Wang,  Z. and Feng, B.},
  year = {2024},
  month = aug,
  pages = {10237–10243}
}

@article{Wannier90,
  title={Wannier90 as a community code: new features and applications},
  author={Pizzi, Giovanni and Vitale, Valerio and Arita, Ryotaro and Bl{\"u}gel, Stefan and Freimuth, Frank and G{\'e}ranton, Guillaume and Gibertini, Marco and Gresch, Dominik and Johnson, Charles and Koretsune, Takashi and others},
  journal={Journal of Physics: Condensed Matter},
  url = {http://dx.doi.org/10.1088/1361-648X/ab51ff},
  DOI = {10.1088/1361-648x/ab51ff},
  volume={32},
  number={16},
  pages={165902},
  year={2020},
  publisher={IOP Publishing}
}

\end{document}